\documentclass[11pt,graphicx,amsmath]{article}
\usepackage{amsmath}
\usepackage{graphicx}
\usepackage{bm}

\title{ Cosmic age, Statefinder and $Om$ diagnostics
          in the decaying vacuum cosmology }

\author{\small   M.L. Tong\thanks{mltong@mail.ustc.edu.cn}\ , Y. Zhang\\
        \small Key Laboratory of Galactic and Cosmological Research\\
        \small Center for Astrophysics \\
       \small University of Science and Technology of China \\
       \small Hefei, Anhui, 230026, China }
 \date{}

\begin{document}
\maketitle

\def\be{\begin{equation}}
\def\ee{\end{equation}}
\def\ba{\begin{eqnarray}}
\def\ea{\end{eqnarray}}
\def\nn{\nonumber}

 \baselineskip=19truept

 \Large

\begin{center}
\Large  Abstract
\end{center}

\begin{quote}
\Large
As an extension of $\Lambda$CDM,
the  decaying vacuum model (DV) describes the dark energy
as a varying vacuum whose energy density
 decays linearly with the Hubble parameter in
 the late-times, $\rho_\Lambda(t) \propto H(t)$,
and produces the matter component. We examine the high-$z$ cosmic
age problem in the DV model, and compare it with $\Lambda$CDM and
the Yang-Mills condensate (YMC) dark energy model. Without employing
a dynamical scalar field for dark energy, these three models share a
similar behavior of late-time evolution. It is found that the DV
model, like YMC, can accommodate the high-$z$ quasar APM 08279+5255,
thus greatly alleviates the high-$z$ cosmic age problem. We also
calculate the Statefinder $(r,s)$ and the {\it Om} diagnostics in
the model. It is found that the evolutionary trajectories of $r(z)$
and $s(z)$ in the DV model are similar to those in the kinessence
model, but are distinguished from those in $\Lambda$CDM and YMC. The
${\it Om}(z)$ in  DV  has a negative slope and its height depends on
the matter fraction, while YMC has a rather flat ${\it Om}(z)$,
whose magnitude depends sensitively on the coupling.

\end{quote}

\newpage

\section{ Introduction}
The observations of the luminosity-redshift relation $d_L(z)$
of type Ia supernovae have indicated that the
expansion of the universe is accelerating \cite{Riess}.
Within the framework of general relativity,
this would imply that there exists a cosmic dark energy (DE)
as a major component in the Universe that drives the acceleration.
Observations of
cosmic microwave background (CMB) \cite{Bennett,DNSpergel,Dunkley},
and large scale structure \cite{Bahcall},
especially baryon acoustic oscillations \cite{Eisenstein},
weak gravitational lensing \cite{Jarvis},
and X-ray clusters \cite{Allen},
have also shown that the universe
is spatially flat with $\sim75\%$ of DE and
$\sim25\%$ of matter.
Besides  the cosmological constant model ($\Lambda$CDM) \cite{Weinberg},
a number of models of DE have been proposed
as the potential source for the acceleration.
Among them there are various scalar models \cite{Wetterich},
vector models, such as the Yang-Mills condensate  (YMC) \cite{Zhang1, Xia},
and modified gravity \cite{Carroll}, and so on.

As an extension to $\Lambda$CDM,
another model  proposed recently is the decaying vacuum (DV) dark energy model
without referring to a specific dynamic field \cite{Borges1,Carneiro}.
In this model, the DE  described by the varying vacuum
is decaying and producing  the matter as the universe expands,
and the equation of state (EOS) of the vacuum is
a constant value $w_\Lambda=p_\Lambda(t)/\rho_\Lambda(t)=-1$,
the same as that in $\Lambda$CDM.
One interesting feature of the DV model is that
its late-time dynamics
is similar to $\Lambda$CDM, and to the YMC dark energy models \cite{Zhang1, Xia}.
The DV model has been tested by observational data of supernova Ia,
yielding a constraint on the present matter density contrast and the Hubble parameter:
$0.27\le \Omega_{m0}\le 0.37$
 and $0.68\le h\le 0.72$ (at $2\sigma$) \cite{Carneiro2}.
A joint statistical analysis of SN Ia,
the baryonic acoustic oscillation,
and the cosmic microwave background anisotropies  (CMB),
gives the constraint
 $\Omega_{m0}\le 0.36\pm 0.01$  and $h=  0.69\pm 0.01$ \cite{Carneiro3}.
The similar joint statistical analysis has also been made,
with a much larger sample of SN Ia \cite{Riess2},
yielding $\Omega_{m0}= 0.289^{+0.013}_{-0.014}$
and $h=  0.634 \pm 0.004$ for the YMC model of 3-loop corrections,
and $\Omega_{m0}= 0.283$
and $h=  0.638$ for $\Lambda$CDM.
The $\chi^2$ statistics has shown that YMC performs slightly
better than $\Lambda$CDM  \cite{Wang1}.
In this paper,
we will confront the DV model with the high-$z$ cosmic age problem,
and examine it by
the Statefinder $(r,s)$ and the {\it Om} diagnostics,
aiming at differentiating it from $\Lambda$CDM, YMC, and several other DE models.

The so-called ``high-$z$ cosmic age problem'' is related to the
quasar APM 08279+5255, whose age $t_{qua}=(2.0-3.0)$ Gyr at the
redshift $z=3.91$,
as evaluated by its chemical evolution
\cite{Hasinger,Friaca}.
The problem has been examined in several DE
models \cite{Friaca,Wang,Alcaniz}.
As is known, the $\Lambda$CDM
model has difficulty in accommodating the high-$z$ quasar.
Ref.\cite{Wang} has shown that the introduction of a coupling between
DE and matter can generally increase the high-$z$
cosmic age.
As an example, it has been found that
the YMC dark energy coupled with the
matter greatly alleviates the problem.
Since the DV scenario is an
interacting DE model, it is also expected to have a longer
high-$z$ cosmic age.

To distinguish the DV model from other DE models,
the Statefinder diagnosis $(r,s)$ can be used \cite{Sahni},
which has been applied to
several other DE models \cite{Gorini,Mltong,Tong}.
As a complementary to  $(r,s)$,
a new diagnostic called {\it Om}
has been recently proposed \cite{Sahniaa},
which helps to distinguish the present matter density contrast $\Omega_{m0}$
in different models more effectively.
We will explore
both  $(r,s)$ and {\it Om} diagnostics  for the DV model,
and compare it with  $\Lambda$CDM and  YMC.
In the following, we  use the  unit $8\pi G=c=1$.

\section{  The decaying vacuum model}

In a spatially flat ($\Omega_{\Lambda0}+\Omega_{m0}=1$)
Robertson-Walker (RW) space-time, the Friedmann equation is
\be\label{fridmann}
\rho_T=3H^2,
\ee
with $H=\dot{a}/a$ being the Hubble rate of expansion,
where an overdot means taking derivative with respect
to the cosmic time $t$.
The energy conservation equation is given by
\be\label{totalconserve}
\dot{\rho}_T+3H(\rho_T+p_T)=0,
\ee
where $\rho_T$ and and $p_T$
  are the total energy and pressure, respectively.
For the present epoch, the universe consists of
the matter (baryons and dark matter) and the DE,
and the radiation can
be ignored.
So $\rho_T=\rho_m+\rho_\Lambda$ and $p_T=p_\Lambda$.
Moreover, in the DV model,
 $\rho_\Lambda =\Lambda(t)$
and  $p_\Lambda=-  \Lambda(t)$.
Thus Eq. (\ref{totalconserve}) can be written as
 \be \label{matter}
\dot{\rho}_m+3H\rho_m=-\dot{\Lambda},
\ee
showing that
the  decaying vacuum density $\Lambda(t)$
plays the role of a source of matter production.
If $\Lambda$ is a  constant,
Eq.(\ref{matter}) reduces to
the continuity equation for matter in $\Lambda$CDM model.
To proceed, one takes the so-called
late-time ansatz \cite{Borges1,Carneiro}
\be \label{ansatz}
\Lambda(t)=\sigma H(t)
\ee
where $\sigma$ is a constant.
Then one gets the following solutions \cite{Borges1}
\ba
\label{a}
&& a (t)=C(e^{{\sigma t}/{2}}-1)^{2/3}\\
 \label{rhomatter}
&&\rho_m(t)=\frac{\sigma^2C^3}{3a^3}
    +\frac{\sigma^2C^{{ 3}/{2}}}{3a^{{ 3}/{2}}},
\\\label{lambdamatter}
&&\Lambda(t)=\frac{\sigma^2}{3}
    +\frac{\sigma^2C^{{ 3}/{2}}}{3a^{{ 3}/{2}}},
\ea
with  $C$ being an integration constant.
If the normalization of the present scale factor $a(t_0)=1$ is taken,
 one can check from Eqs.(\ref{rhomatter}) and
(\ref{lambdamatter}) that
\be
C = \left(\frac{\Omega_{m0}}{1-\Omega_{m0}}\right)^{2/3},
\ee
and $\sigma = 3\Omega_\Lambda H_0$,
where $H_0=100\, h$ km s$^{-1}$ Mpc$^{-1}$
is the present Hubble constant
and $h$ is the Hubble parameter.
The
first terms in Eqs. (\ref{rhomatter}) and (\ref{lambdamatter})
give the usual scaling of
the matter and the vacuum, respectively,
while the second terms describe the matter production
caused by the decaying vacuum.
It is these interacting terms
between the vacuum and the matter \cite{Wang}
that will be likely to increase the high-$z$ cosmic age
in the DV model.
Eqs.(\ref{a}), (\ref{rhomatter}) and  (\ref{lambdamatter})
give an smooth transition between the matter with $a\ll1$
and vacuum epoch with $a\gg1$.
In the following,
we will examine the high-$z$ cosmic age problem
and the Statefinder and $\it Om$ diagnostics in  the DV model.

\section{  High-$z$ cosmic age problem}

As is known, introducing the cosmological constant $\Lambda $ can
increase the predicted cosmic age. Similarly, the presence of
$\Lambda(t)$ in the DV model also increases the present cosmic age
with $t_0\sim15$ Gyr \cite{Carneiro3}, in concordance with the
current estimated one $t_0\sim 14$ Gyr from a combination of data of
Wilkinson Microwave Anisotropy Probe (WMAP) and Sloan Digital Sky
Survey (SDSS) \cite{Tegmark}. However, the high-$z$ cosmic age
problem is more difficult to solve. For the quasar APM 08279+5255,
the estimated age $t_{qua}=(2.0-3.0)$ Gyr at $z=3.91$ has been based
upon the observed high  Fe/O abundance ratio $\sim 3$ from the X-ray
data \cite{Hasinger} and upon the chemodynamical modeling for the
evolution \cite{Hermann}. For instance, in a detailed investigation,
Ref.\cite{Friaca} gives an age $t_{qua}= 2.1$ Gyr at $z=3.91$. For a
cosmological model to solve the high-$z$ problem, its predicted
cosmic age at $z=3.91$ should be older than the lower limit of the
observed age of $2.0$ Gyr.

Given a cosmological model,
the cosmic age $t(z)$ as a function of redshift $z$
is written as
\be\label{age1}
t(z)=\int_z^\infty\frac{d\tilde{z}}{(1+\tilde{z}) H(\tilde{z})},
\ee
where $H(z)$ is the Hubble parameter in terms of redshift $z$.
For the DV model,
using Eqs. (\ref{fridmann})  and (\ref{a})-(\ref{lambdamatter}),
one obtains
\be\label{hubblez}
 H(z)=H_0[1-\Omega_{m0}+\Omega_{m0}(1+z)^{3/2}].
\ee
For comparison,
we also list
\be\label{hubblel}
H(z) =H_0[1-\Omega_{m0}+\Omega_{m0}(1+z)^3]^{1/2}
\ee
for the $\Lambda$CDM model, and
\be \label{hubbleYM}
H(z)  =H_0 [(1-\Omega_{m0})\frac{\rho_y(z)}{\rho_{y0}}
         +\Omega_{m0}  \frac{\rho_m(z)}{\rho_{m0}}   ]^{1/2}
\ee
for the YMC DE model \cite{Zhang1,Xia,Wang1,Wang}.
In  Eq.(\ref{hubbleYM}), $\rho_y(z)$ and $\rho_m(z)$
are the energy density of the YMC
and of the matter, respectively,
which depend implicitly on the coupling between the YMC and the matter.
Here we will consider the case of  2-loop YMC
decaying into the matter
at a constant decay rate  $\Gamma  \sim H_0$  \cite{Xia}.

Substituting Eqs.(\ref{hubblez}), (\ref{hubblel}), and
(\ref{hubbleYM}), respectively, into Eq.(\ref{age1}), yields the
cosmic age $t(z)$ predicted by the DV, the $\Lambda$CDM, and the YMC
model. Since  $H(z)$ depends explicitly the parameters $h$ and
$\Omega_{m0}$, so does the cosmic age $t(z)$ given by
Eq.(\ref{age1}). With other parameters fixed, a larger $h$ tends to
yield a smaller $t(z)$, and so does a larger $\Omega_{m0}$.
Currently, the value of $h$ is still in debate even after a number
of experiments and observations of various kinds in the past.

For instance, the final result from the Hubble Space Telescope (HST)
Key Project \cite{Freedman} is $h=0.72\pm0.08$, and Riess et al.
recently give $h=0.742\pm0.036$ \cite{Riess8}. Whereas Sandage et
al. \cite{Sandage} advocate the global value $h=0.623\pm0.063$ from
HST using SN Ia, calibrated with Cepheid variables in nearby
galaxies. From the observational data on CMB anisotropies, WMAP3
\cite{DNSpergel} and WMAP5  \cite{Dunkley} give $h=0.732\pm0.031$
and  $h=0.701\pm0.013$, respectively. On the other hand, the value
of $\Omega_{m0}$ is less uncertain, and has been determined by WMAP3
\cite{DNSpergel}, WMAP5  \cite{Dunkley},
 and SDSS \cite{Tegmark},
to be $0.24$, $0.25$, and $0.24$, respectively.

\begin{figure}
\centerline{\includegraphics[width=12cm]{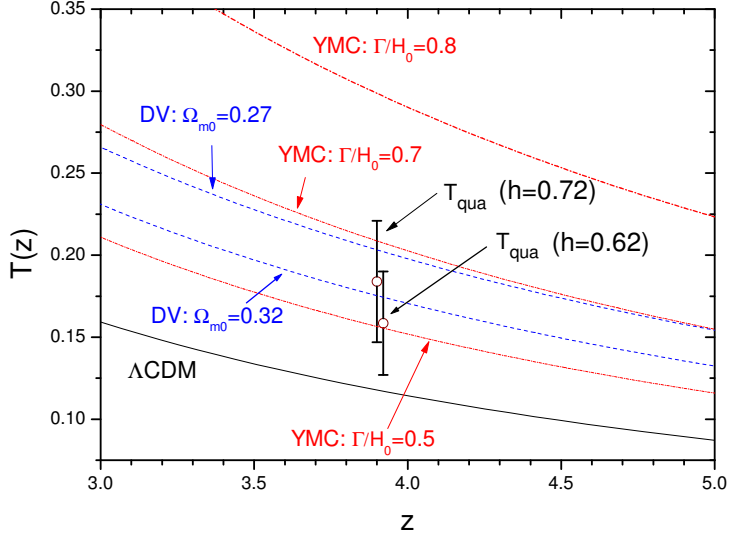}}
 \caption{\label{age}\small
 The dimensionless cosmic age $T(z)\equiv H_0t(z)$
 are plotted for three models.
For the APM 08279+5255 with age  $t_{qua}= (2.0-3.0)$ Gyr,
the two vertical line segments at $z\sim3.91$
indicate $T_{qua} = (0.147,0.221)$  for $h=0.72$,
 $T_{qua} = (0.127,0.190)$  for $h=0.62$.
Here $\Omega_{m0}=0.27$ is taken for $\Lambda$CDM and YMC.
 }
\end{figure}

It is convenient to use
a dimensionless cosmic age $T(z)\equiv H_0t(z)$,
which depends on $\Omega_{m0}$ only.
For three DE models with  $ \Omega_{m0}=0.27$:
DV, $\Lambda$CDM, and 2-loop YMC \cite{Xia},
Fig.\ref{age} shows the respective $T(z)$.
Besides,  $T(z)$ of  DV model with $\Omega_{m0}=0.32$
is also plotted.
The dimensionless  age $T_{qua}\equiv H_0t_{qua}$
of  the quasar APM 08279+5255
depends on the parameter  $h$.
With  $t_{qua}= (2.0-3.0)$ Gyr,
$T_{qua}$ has a range of $(0.147,0.221)$ for $h=0.72$
and of $(0.127,0.190)$  for $h=0.62$ at $z\sim3.91$.
These are represented by the vertical line segments in Fig. \ref{age}.

The Universe can not be younger than its constituents at any redshift.
Specifically, in regard to the quasar APM 08279+5255,
the cosmic age  at $z=3.91$ predicted in
a cosmological model must be greater than the quasar age,
\be  \label{constrain}
t(3.91)\geq    t_{\rm qua}, \,\,\,\,\,
    i.e., \,\,\,\,   T(3.91)\geq    T_{\rm qua}.
\ee
Fig.\ref{age}  shows that,
 for either $h=0.72$ or $0.62$,
the cosmic age $t(z)$ predicted by  DV lies safely above
the lower limit $t_{qua}=2$Gyr at $z=3.91$.
For  $h=0.62$ and $\Omega_{m0}=0.27$,
$t(z)$ is even higher than the upper limit $t_{qua}=3$Gyr at $z=3.91$.
Thus we can say that
the DV model greatly alleviates the high-$z$ cosmic age problem.
Fig.\ref{age} also shows that
a low $\Omega_{m0}$ yields a higher $t(z)$.
The reason that  DV has a higher cosmic age
is that the model intrinsically has an interaction
between the DE and the matter.
These interaction terms are fully determined
by $H_0$ and $\Omega_{m0}$, and are not adjustable.
On the other hand, for YMC
the interaction is realized by the decay rate $\Gamma$
as a  model parameter,
and a higher rate $\Gamma$ will yield a longer comic age.
For instance,
the YMC with $\Gamma=0.7H_0$ performs a bit better than the DV,
and has an age $T(z)$ near the upper limit $T_{qua}$ even  for $h=0.72$.
As shown in Fig.\ref{age}, the $\Lambda$CDM
cannot accommodate the quasar even for $h=0.62$.

\section{  Statefinder and {\it Om} diagnostics in the DV model}

\begin{figure}
\centerline{\includegraphics[width=12cm]{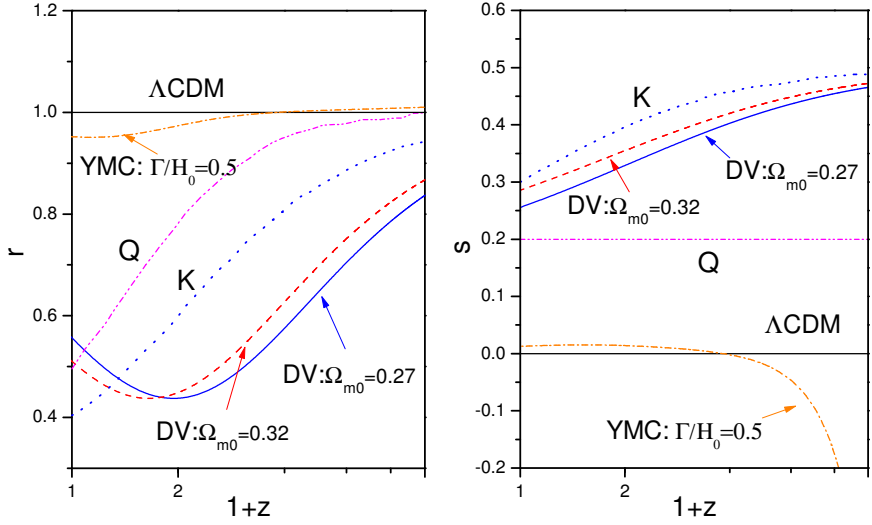}}
 \caption{\label{rsevolution}\small
 $r(z)$ and $s(z)$ in the DV model for two cases
 with $\Omega_{m0}=0.27$ and $0.32$.
 For comparison, the results are also plotted
 for other DE models including
  the $\Lambda$CDM,
  the coupling YMC with a decay rate $\Gamma=0.5 H_0$,
  quiessence (Q), and kinessence (K) models \cite{Sahni}.
}
 \end{figure}

The Statefinder diagnostic pair $\{r,s\}$ are defined as \cite{Sahni}
\be\label{RS}
r\equiv\frac{\stackrel{\dots}{a}}{aH^3}~,
~~~~s\equiv\frac{r-1}{3(q-1/2)}~,
\ee
where $q=\frac{\ddot{a}}{aH^2}$ is the deceleration parameter.
These parameters can be also expressed
in terms of  $\rho_T$
and  $p_T$  as follows
 \be\label{rs}
 q=\frac{1}{2}(1+\frac{3p_T}{\rho_T}),~~~
r=1+\frac{9(\rho_T+p_T)\dot{p}_T}{2\rho_T\dot{\rho}_T},~~~
s=\frac{\rho_T+p_T}{p_T}\frac{\dot{p}_T}{\dot{\rho}_T}.
 \ee
For the DV model, one has
\ba\label{statefinder}
&&q=\frac{3}{2}\Omega_m-1,\\
&&r=1-\frac{9}{4}\Omega_m(1-\Omega_m)    ,\\\label{s}
&&s=\frac{1}{2}\Omega_m,
\ea
where $\Omega_m=\rho_m(t)/\rho_T(t)$
is the time dependent matter fraction.
By Eqs.  (\ref{a})$-$(\ref{lambdamatter})
it can be written as
 \be\label{omegam}
 \Omega_m=e^{-\sigma t/2}
    =\frac{\Omega_{m0}(1+z)^{3/2}}{[1-\Omega_{m0}+\Omega_{m0}(1+z)^{3/2}]}.
  \ee
Substituting it  into Eqs. (\ref{statefinder})$-$(\ref{s}) yields
\ba\label{statefinderq}
&&q=\frac{3\Omega_{m0}(1+z)^{3/2}}{2[1-\Omega_{m0}+\Omega_{m0}(1+z)^{3/2}]}-1,\\
&&r=1-\frac{9\Omega_{m0}(1-\Omega_{m0})(1+z)^{3/2}}
{4[1-\Omega_{m0}+\Omega_{m0}(1+z)^{3/2}]^2},\\\label{statefinders}
&&s=\frac{\Omega_{m0}(1+z)^{3/2}}{2[1-\Omega_{m0}+\Omega_{m0}(1+z)^{3/2}]},
\ea
which depend on the parameter $\Omega_{m0}$.
We plot the parameters $r(z)$ and $s(z)$ in Fig.\ref{rsevolution}
for the DV model with $\Omega_{m0}=0.27$
and with $\Omega_{m0}=0.32$, respectively.
For comparison,
$r(z)$ and $s(z)$ in $\Lambda$CDM,  YMC,
quiessence,  kinessence  models \cite{Sahni}
are also plotted there.
The left panel of Fig.\ref{rsevolution} shows
that $r(z)$ in the DV model has a minimum at $z\sim1$,
a feature drastically distinguished from other models.
Since this feature appears around  $z\sim 1$,
where observational data are easier to obtain,
direct confrontations of the DV model with observations
is feasible.
Moreover, the right panel of Fig.\ref{rsevolution} shows  that
the behavior of $s(z)$ in the DV model is similar to
that in the kinessence model,
different from other models.
In particular,
it should be pointed out that although DV and $\Lambda$CDM
have the same EOS $w=-1$,
their Statefinder pair $(r,s)$ are effectively distinguished.
On the other hand, amongst the various models plotted,
YMC has a Statefinder $r(z),s(z)$
that is most close to that of  $\Lambda$CDM
with $(r,s)=(1,0)$.

\begin{figure}
\centerline{\includegraphics[width=12cm]{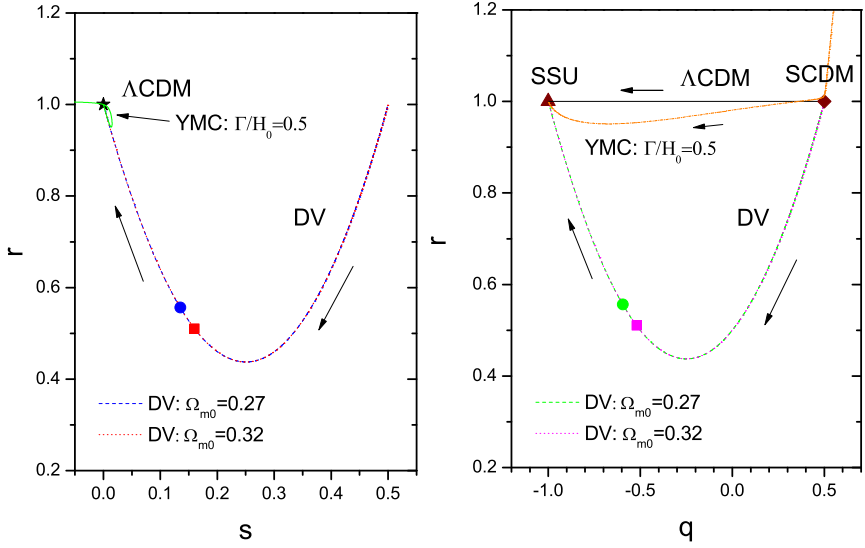}}
 \caption{\label{rqs}\small
 The trajectories of $(r,s)$ and of $(r,q)$ in the DV model for
 two cases: $\Omega_{m0}=0.27$ and $\Omega_{m0}=0.32$.
 The
filled circles and squares stand for the cases of $\Omega_{m0}=0.27$
and  $\Omega_{m0}=0.32$,
 respectively.
 The star in the $r-s$ plane means the fixed point of
 $\Lambda$CDM.
 The diamond and the triangle
 in the $r-q$ plane denote the fixed
points of the standard cold dark matter (SCDM) model   and of the
steady state Universe (SSU) \cite{Tong}, respectively.
 The arrows along the curves are the evolution direction.}
 \end{figure}

The trajectories of ($r,s$) and of ($r,q$)
can also help to differentiate various DE models.
The results of DV, YMC, $\Lambda$CDM are plotted in Fig.\ref{rqs}.
These two trajectories distinguish the DV from
$\Lambda$CDM and YMC very effectively.
The shape of the  trajectories of $(r,s)$ of DV
are quite similar to those of
kinessence model given by Fig.1 in Ref.\cite{Sahni},
which, nevertheless, has a time dependent EOS $w\neq $constant.
Also notice that the shape of the curve $s-r$ is quite similar
to that of the curve $q-r$.
This is easy to understand since
$q$ and $s$ are linearly related by $q=3s-1$,
according to Eqs. (\ref{statefinderq}) and (\ref{statefinders}).
From Fig. \ref{rsevolution} and Fig. \ref{rqs}
we notice that the Statefinder $(r,s)$ for DV
is not very sensitive to the matter fraction $\Omega_{m0}$.
For instance,  for $\Omega_{m0}=0.27$ and $\Omega_{m0}=0.32$,
the left panel in Fig. \ref{rqs}
shows almost overlapping trajectories $(r,s)$,
and the right panel shows almost overlapping trajectories $(r,q)$.
This kind of degeneracy can be broken to certain extent
by the {\it Om} diagnostic as follows.

\begin{figure}
\centerline{\includegraphics[width=12cm]{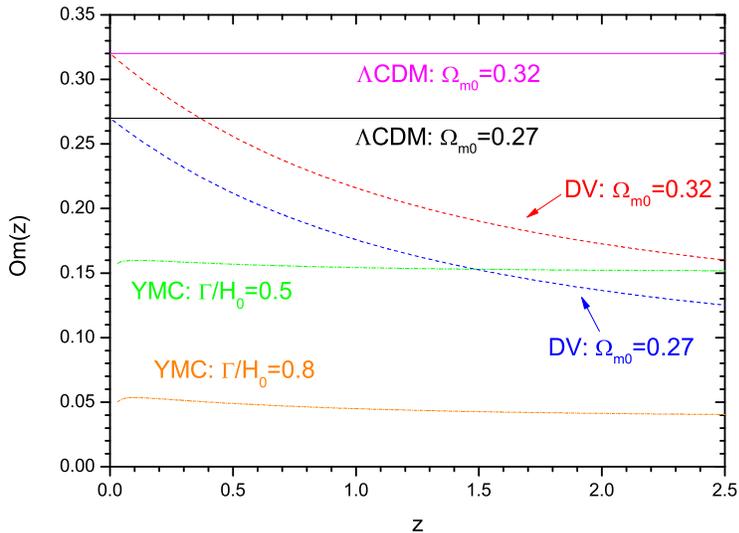}}
 \caption{\label{om}\small
${\it Om}(z)$  in  DV has a negative slope
and its height depends sensitively on $\Omega_{m0}$.
${\it Om}(z)$ in  YMC is rather flat and
its height depends on the coupling $\Gamma$. }
 \end{figure}

Now we turn to the {\it Om} diagnostic
defined as \cite{Sahniaa}
 \be
  Om(x)=\frac{h^2(x)-1}{x^3-1},
  \ee
where $x\equiv (1+z)$ and $h(x)\equiv H(x)/H_0$.
Thus $Om$ involves only the first derivative of the scale factor
through the Hubble  parameter
and is easier to reconstruct from observational data.
For $\Lambda$CDM  with Eq.(\ref{hubblel}), it is simply
 \be Om(x)=\Omega_{m0},
\ee
independent of  redshift.
For  DV with Eq. (\ref{hubblez}),  one has
\be
Om(x)=\frac{(1-\Omega_{m0}+\Omega_{m0} x^{3/2} )^2-1}{x^3-1}.
\ee
For YMC,  one has
\be
Om(x)=\frac{ [(1-\Omega_{m0})\frac{\rho_y(x)}{\rho_{y0}}
         +\Omega_{m0} \frac{\rho_m(x)}{\rho_{m0}}    ]^2-1}{x^3-1}.
\ee
Fig. \ref{om} shows  ${\it Om}(z)$ for $z \le 2.5$
in DV and  YMC,
both being distinguished from $\Lambda$CDM.
  ${\it Om}(z)$ in DV has a negative slope,
similar to that quintessence models. Moreover, the {\it Om}
diagnostic distinguishes  effectively the two cases
$\Omega_{m0}=0.27$ and $\Omega_{m0}=0.32$ of the DV model. This is
an advantage of $\it Om$ over the Statefinder diagnostic. Besides,
although ${\it Om}(z)$ are quite flat in both  YMC and $\Lambda$CDM,
${\it Om}(z)$ in YMC lies far below that of $\Lambda$CDM. So they
are also differentiated more easily by the {\it Om} diagnostic than
by the Statefinder diagnostic \cite{Mltong,Wang1}. Furthermore, for
a fixed $\Omega_m$, the  Om(z) in the DV model approaches
      that in the $\Lambda$CDM model as $z\rightarrow0$,
but the flat ${\it Om}(z)$ in  YMC does not have that property.

\section{Summary}

We have demonstrated that, the DV dark energy model
with a low value of $\Omega_{m0}$
can marginally accommodate the high-$z$ quasar APM 08279+5255,
if the Hubble parameter  $h= 0.72$ or smaller.
This is due to the fact that
the decaying $\Lambda(t)$
intrinsically induces an interaction with the matter component,
thus increasing the cosmic age.
To solve the age problem completely,
a lower value $\Omega_{m0}\le 0.27$ would be required.
But observational data of SN Ia has already put
a constraint  $0.27\le  \Omega_{m0} \le 0.36$ on the DV,
so one can only say that
the DV alleviates  the high-$z$ cosmic age problem.
In comparison,  the interacting YMC model
can solve the problem by increasing the decay rate $\Gamma$.

To examine DV further,
we have also calculated the Statefinder diagnostic $(r,s)$ in the model,
and have made comparison
with $\Lambda$CDM, YMC, quiessence, and kinessence.
It has been found that
$(r,s)$ effectively distinguishes DV  from
$\Lambda$CDM,
even though the two models have the same EOS $w=-1$.
Besides, $(r,s)$ in DV is also different from
those in YMC,  quiessence, and kinessence.
The trajectories of the
 pair $(r,s)$ and $(r,q)$ have a similar shape in  DV
because of a linear relation between $s$ and $q$.

While the Statefinder $(r,s)$ for DV
is not very sensitive to  $\Omega_{m0}$,
the {\it Om} diagnostic has been employed to break this degeneracy.
It has been shown that
the  ${\it Om}(z)$ in DV has a negative slope
and its height depends on the matter fraction,
while YMC has a rather flat ${\it Om}(z)$,
which depends sensitively on the magnitude of the coupling.
Therefore, the ${\it Om}$ diagnostic
will be quite  effective in distinguishing
$\Lambda$CDM, DV, and YMC models
from observational data of relatively low redshifts.

\

{ACKNOWLEDGMENT}:
M.L. Tong's work has been partially supported
by Graduate Student Research Funding from USTC.
Y.Zhang's research work has been supported by the CNSF
No.10773009, SRFDP, and CAS.

\end{document}